\documentclass[10pt,leqno]{amsart}

\usepackage{indentfirst,csquotes}

\usepackage{xcolor,paralist,hyperref,fancyhdr,etoolbox}

\hypersetup{ colorlinks=true, linkcolor=black, filecolor=black, urlcolor=black }

\usepackage{breqn}
\usepackage[T1]{fontenc}
\usepackage[utf8]{inputenc}
\DeclareUnicodeCharacter{03BE}{\ensuremath{\xi}}
\DeclareUnicodeCharacter{03BC}{\ensuremath{\mu}}
\DeclareUnicodeCharacter{03C3}{\ensuremath{\sigma}}

\graphicspath{{plots/}}

\usepackage[numbers, square]{natbib}

\usepackage{hyperref}

\usepackage{gensymb}

\begin{document}

\title{Extending the blended generalized extreme value distribution}
\author[Krakauer]{Nir Y. Krakauer}
\address{Department of Civil Engineering, The City College of New York, New York, NY 10031, USA}
\email{nkrakauer@ccny.cuny.edu}

\begin{abstract}
The generalized extreme value (GEV) distribution is commonly employed to help estimate the likelihood of extreme events in many geophysical and other application areas. The recently proposed blended generalized extreme value (bGEV) distribution modifies the GEV with positive shape parameter to avoid a hard lower bound that complicates fitting and inference. Here, the bGEV is extended to the GEV with negative shape parameter, avoiding a hard upper bound that is unrealistic in many applications. This extended bGEV is shown to improve on the GEV for forecasting heat and sea level extremes based on past data. Software implementing this bGEV and applying it to the example temperature and sea level data is provided.
\end{abstract}

\maketitle

\bigskip

\section{Introduction}

The generalized extreme value (GEV) distribution can be derived as the limit distribution of block maxima (from any distribution for which such a limit exists) for sufficiently large blocks and numerous samples (the Fisher–Tippett–Gnedenko theorem) \cite{haan2006}. However, in practice the GEV distribution is fitted to finite data sets, such as observations of temperature, precipitation, streamflow, and sea level (among many other applications to problems in fields ranging from finance to biology), to estimate the likelihood of future extreme events \cite{jenkinson1955, rypkema2021}. ``Engineering design is primarily concerned with the extremes'', which, in the context of buildings and civil infrastructure, requires accurate assessment of the frequency of the most impactful weather events and how this might change over the design lifetime \cite{cacc15}. In this context, some properties of the GEV distribution complicate inference.

The GEV is a function of 3 parameters, a real shape parameter ξ, real location parameter μ, and positive scale parameter σ. In terms of standardized values $s = \frac{x - μ}{σ}$, its cumulative distribution function can be written as

\begin{equation}
F_{\text{GEV}}(s) = 
\begin{cases}
	\exp\left( -e^{-s}\right) & \text{if } ξ = 0\\
	\exp\left( -\left( 1 + ξs \right)^{-\frac{1}{ξ}} \right) & \text{if } ξ \neq 0 \text{ and } ξs > -1\\
	0 & \text{if } ξ > 0  \text{ and } ξs \leq -1\\
	1 & \text{if } ξ < 0  \text{ and } ξs \leq -1. 
\end{cases}  
\end{equation}

The GEV distribution thus has positive density for all real $x$ only in the Gumbel case, when $ξ = 0$. In the Fréchet case, when $ξ > 0$, there is zero probability that $\frac{x - μ}{σ} \leq -\frac{1}{ξ}$. In the Weibull case, when $ξ < 0$, there is zero probability that $\frac{x - μ}{σ} \geq -\frac{1}{ξ}$. This property does not match well the behavior of finite sets of block maxima derived, for example, from geophysical phenomena, which typically do not have an intrinsic sharp upper or lower bound. While admissible GEV parameters can be found for any collection of block maxima, forecasting with any such distribution may lead to surprises where a new record value had zero probability. As a result, a GEV distribution would be arbitrarily bad when assessed using the expected likelihood of new observations \cite{benedetti10, smith15}. Moreover, some common computational tools for inference with data, such as automatic differentiation variational inference \cite{kucukelbir2017}, require the statistical distributions used to have unbounded support, and so cannot be easily employed with the GEV distribution.

In order to overcome such problems, \cite{castro_camilo2022} suggested a blended generalized extreme value (bGEV) distribution as a blend of the GEV distribution with $ξ > 0$ and the Gumbel distribution (i.e. GEV with $ξ = 0$), where the blended distribution would be identical to the Fréchet case over most of its probability mass but still enjoy unbounded support because its right tail follows that of a Gumbel distribution. Their bGEV cumulative distribution function can be written as

\begin{equation}
\label{eq:bgev}
F_{\text{bGEV}}(s) = 
F_{\text{GEV}}(s)^{p(s)} F_{\text{Gumbel}}(\tilde{s})^{1 - p(s)}.
\end{equation}

$F_{\text{GEV}}(s)$ is the GEV distribution at $s = \frac{x - μ}{σ}$ with the chosen parameters $ξ > 0, μ, σ$. $F_{\text{Gumbel}}(\tilde{s})$ is the Gumbel distribution with location and scale parameters $\tilde{μ}, \tilde{σ}$ used to standardize $\tilde{s} = \frac{x - \tilde{μ}}{\tilde{σ}}$. $\tilde{μ}, \tilde{σ}$ are chosen such that $F_{\text{Gumbel}}(\tilde{s}_a) = F_{\text{GEV}}(s_a) = a$ and $F_{\text{Gumbel}}(\tilde{s}_b) = F_{\text{GEV}}(s_b) = b$ for two user-specified quantiles $0 < a < b < 1$. $p(s)$ is a function that equals 0 for $s \leq s_a$, 1 for $s \geq s_b$, and, for intermediate values $s_a < s < s_b,$ is equal to the cumulative distribution function of the beta distribution $F_{\text{Beta}}(\frac{s - s_a}{s_b-s_a}; \alpha, \beta)$ with user-specified positive shape parameters $\alpha, \beta$. \cite{castro_camilo2022} recommend $a = 0.05, b = 0.2, \alpha = \beta = 5$, and also provide more details on this new distribution.

This bGEV distribution retains the fatter right tail of the GEV Fréchet case, while having the Gumbel distribution's unbounded left tail. \cite{castro_camilo2022} show an example application to monthly maximum air pollutant concentrations, and the bGEV distribution has also been applied to annual maximum precipitation intensities \cite{vandeskog2022} and to currency exchange rates \cite{metwane2023}. However, whereas precipitation extremes typically have tail behavior consistent with positive ξ \cite{koutsoyiannis2004}, other types of meteorological extremes such as temperature \cite{frias12, rai2024} and sea level \cite{mendez2007} instead have probability distributions more consistent with negative ξ. Fitting samples to the GEV distribution with negative ξ means that the fitted distribution has a hard upper bound and therefore critically underestimates the occurrence of, e.g., new heat extremes \cite{zeder2023}, as well as posing computational challenges as outlined above.

The bGEV probability density function (PDF), obtained by differentiating Equation \ref{eq:bgev} with respect to $x$, is more complicated for the intermediate or mixing region where $0 < p(s) < 1$, and is given in the Appendix to \cite{castro_camilo2022}, along with its first and second derivatives with respect to $x$.

To address this problem, this note proposes a simple extension of the bGEV to the $ξ < 0$ case, and applies it to example temperature and sea level data. The principle of blending the GEV with the Gumbel distribution at its hard limit so that the support is unbounded is the same, and the details are closely analogous to those derived by \cite{castro_camilo2022} for $ξ > 0$, except for that the blending takes place near the upper rather than the lower tail.

Precisely, for negative ξ, the proposed bGEV has cumulative distribution function 

\begin{equation}
F_{\text{bGEV}}(s) = 
F_{\text{GEV}}(s)^{p(s)} F_{\text{Gumbel}}(\tilde{s})^{1 - p(s)}.
\end{equation}

$F_{\text{GEV}}(s)$ is the GEV distribution at $s = \frac{x - μ}{σ}$ with the chosen parameters $ξ < 0, μ, σ$. $F_{\text{Gumbel}}(\tilde{s})$ is the Gumbel distribution with location and scale parameters $\tilde{μ}, \tilde{σ}$ used to standardize $\tilde{s} = \frac{x - \tilde{μ}}{\tilde{σ}}$. $\tilde{μ}, \tilde{σ}$ are chosen such that $F_{\text{Gumbel}}(\tilde{s}_a) = F_{\text{GEV}}(s_a) = a$ and $F_{\text{Gumbel}}(\tilde{s}_b) = F_{\text{GEV}}(s_b) = b$ for two user-specified quantiles $1 > a > b > 0$. In this case $a, b$ will typically be closer to 1 so modification of the Weibull GEV distribution is concentrated near its problematic upper bound, as opposed to the $ξ > 0$ case where $a, b$ will typically be chosen to be closer to 0. $p(s)$ is a function that equals 0 for $s \geq s_a$, 1 for $s \leq s_b$, and, for intermediate values $s_b < s < s_a,$ is equal to the cumulative distribution function of the beta distribution $F_{\text{beta}}(\frac{s - s_a}{s_b-s_a}; \alpha, \beta)$ with user-specified positive shape parameters $\alpha, \beta$. 

The PDF of the proposed bGEV is also formally the same as that given by \cite{castro_camilo2022} in the mixing region, while being equal to the GEV distribution PDF in the lower part where $p(s) = 1$ and to the Gumbel distribution PDF in the upper part where $p(s) = 0$. Finally, the limit of the bGEV distribution as ξ approaches 0 from either direction is simply the Gumbel distribution. Unlike for the GEV, the mean and variance of the bGEV do not have simple analytic forms, although they can be computed numerically by integrating over the PDF. In practice, these bGEV properties are often close to those of the corresponding GEV, although higher (for negative ξ) since the bGEV is the same as the GEV at the lower quantiles but is unbounded above. For example, if $ξ = -0.3, μ = 0, σ = 1$, the GEV mean and variance (to 8 decimal places) are 0.34176435 and 0.97846332 respectively, while the bGEV values with $ξ = -0.3, μ = 0, σ = 1, a = 0.95, b = 0.8, \alpha = \beta = 5$ are 0.35018832 and 1.02559938. The CDF of the bGEV also does not have a simple analytic inverse in the mixing region where $F_{\text{bGEV}}(s(x))$ is between $a$ and $b$, but quantiles in that range can be found numerically; for example, for $ξ = -0.3, μ = 0, σ = 1, a = 0.95, b = 0.8, \alpha = \beta = 5$, the 90th percentile of the bGEV is 1.61258469. 

%k = -0.3
%ma = (gamma(1 - k) - 1) / k
%mb = integral (@(x) x.* bgevpdf(x, k, 1, 0), -Inf, Inf, "RelTol", 1E-12);
%va = (gamma(1-2*k) - gamma(1-k)^2) / (k^2)
%vb = integral (@(x) (x-mb).^2.* bgevpdf(x, k, 1, 0), -Inf, Inf, "RelTol", 1E-12);
%bgevinv(0.9, k, 1, 0)

\section{Example simulations}

Simulation case A involved drawing samples from a GEV distribution with $ξ=-0.2, μ=0, σ=1$. This distribution has zero probability for $x \geq μ - σ/ξ = 5$. Over 100 realizations, the maximum-likelihood GEV fit to 100 samples drawn from this distribution has, in the median, an upper threshold (with zero probability mass above it) of 4.70, meaning that while the generating distribution does have an upper threshold (of 5), most of the GEV fits have an upper threshold that is too low, so that there is a possibility of a surprise value from this distribution that is above the maximum possible in the fitted distribution. In one realization, the fitted GEV had an upper threshold as low as 2.83, the 98.5 percentile of the generating distribution. By contrast, the bGEV (with $a = 0.91, b = 0.90, \alpha = \beta = 5$) was fitted with negative ξ, but by design had no hard upper threshold, obviating the problem of zero probability mass assigned to part of the generating distribution, although the bGEV could not capture the drop-off of the generating probability density to 0 at $x = 5$.

\begin{figure}[htbp]
	\begin{center}
		\includegraphics[width=12cm]{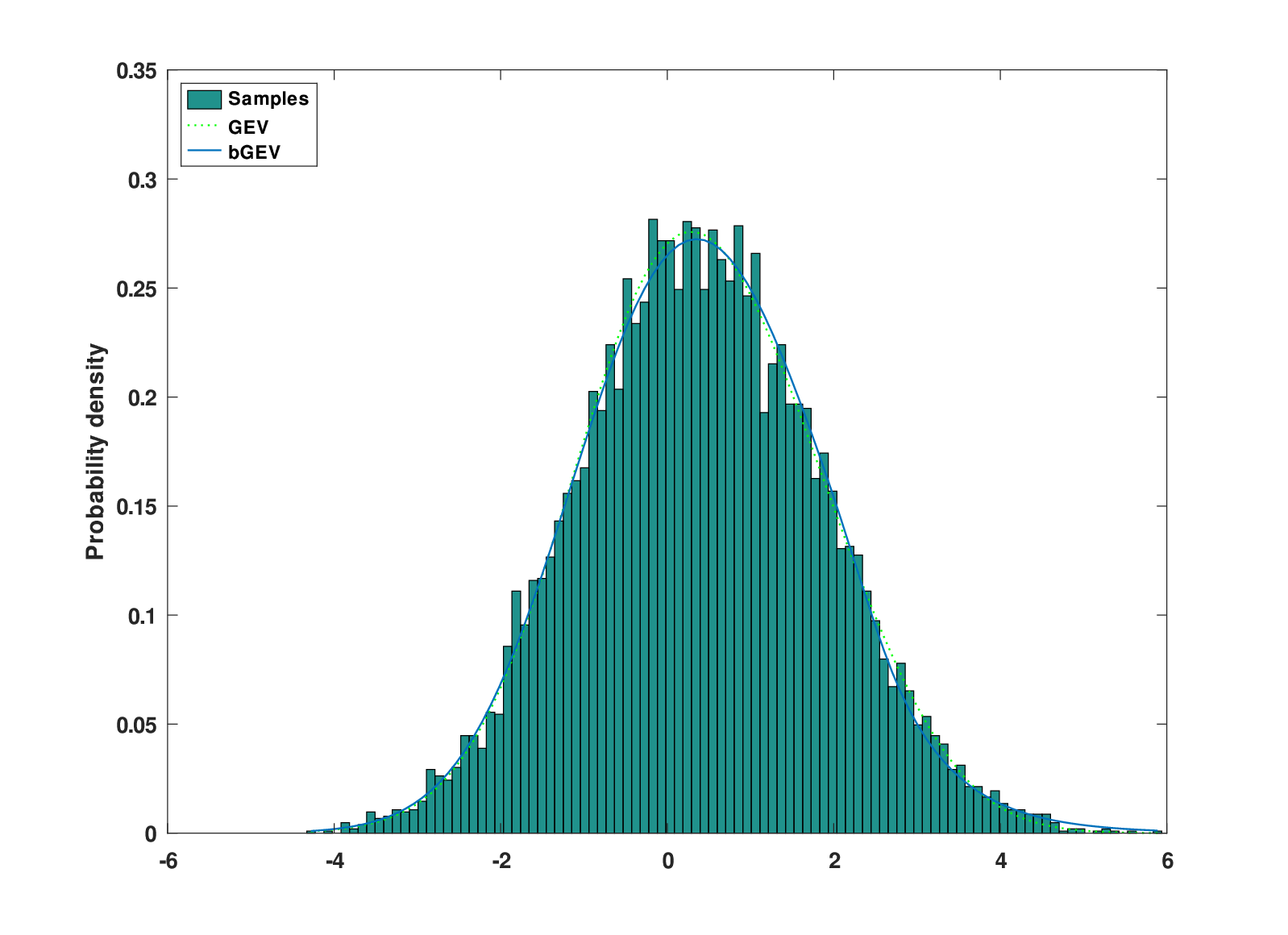}
	\end{center}
	\caption{GEV and bGEV fits to synthetic data (case B). The GEV and bGEV probability density functions shown are averaged over 100 realizations, fit to 100 independent samples each. All 10,000 samples are included in the histogram.}
	\label{fig:fit_caseB} 
\end{figure}

Simulation case B involved drawing samples from a GEV distribution with $ξ=-0.2, μ=0, σ=1$ but with an independent, normally distributed term with mean 0 and standard deviation 1 added to each sample, so that there is no hard threshold anymore; rather, the upper tail falls off like that of a normal distribution. The GEV fitted to 100 samples still indicated a negative shape parameter, giving a median upper threshold of 5.00, which was now qualitatively incorrect since the generating distribution actually had no upper limit. The bGEV's shape, with no upper threshold, was qualitatively correct. (Figure \ref{fig:fit_caseB}) (The bGEV parameters $a, b, \alpha, \beta$ could potentially be adjusted in order to improve quantitative fit to the upper tail of the sampled distribution, if this was important in a real application.)

\section{Example applications}

Recent years have seen long-standing temperature records in many countries broken by unprecedented amounts. The state-of-the-art ERA5 reanalysis \cite{hersbach18, krakauer2023a} shows this pattern well, with most of the globe reaching new temperature records in the last 15 years. As a preliminary evaluation of the ability of GEV and bGEV distributions to fit this type of data, a sample was taken of 100 populated land grid cells spaced at least 10{\degree} apart, sufficient for good global coverage. For each chosen grid cell, annual maximum temperatures for 1940-2023, in units of K, were determined from the ERA5 hourly 2-m air temperature series, resulting in a time series of $n = 84$ block maxima. Such temperature series are frequently fitted using the GEV distribution, as for example reflected in work by the World Weather Attribution collaboration in estimating the anthropogenic contribution to the probability of observed heat extremes occurring \cite{philip2020, oldenborgh2021, philip2022}.

The GEV and bGEV distributions were compared in terms of one-year-ahead probabilistic forecasts, using summed negative log likelihood (NLL) of the actual value in the forecast as the skill metric \cite{du2021}. NLL, also referred to as the ignorance score \cite{roulston02}, is closely related to cross-entropy and Kullback–Leibler divergence, information-theory measures of differences between a given distribution and an imperfect model of it \cite{good52, boer2005, bulinski2021}, with lower NLL corresponding to closer agreement between the probabilistic forecast and the observed value. NLL has previously been used to quantitatively compare different methods for generating probabilistic temperature forecasts \cite{smith15, krakauer15, aizenman16}.

The probabilistic forecasts were generated by fitting, for each length $s$ from 30 up to $n-1$, maximum-likelihood GEV and bGEV distributions for the first $s$ annual maxima, and using those to generate a forecast for year $s+1$ that could be compared to the observed value. (The bGEV was allowed to have a shape parameter of either sign, using the extension outlined above.) NLL was therefore summed over 5400 separate forecasts (54 years for each of 100 locations).

In order to capture the observed global warming trend, the location parameter in the GEV distribution was assumed to be a linear function of the mean annual global temperature (also calculated from ERA5), i.e., for each location, $μ(t) = μ_0 + μ_t(T(t) - \bar{T})$, where $\bar{T}$ is an average historic global temperature. Parameters for the GEV and bGEV -- $μ_0, μ_t, σ, ξ$ -- were estimated for each of the 5400 cases by maximum likelihood. For the bGEV, the beta distribution shape parameters $\alpha, \beta$ were kept at 5. For the cases where the GEV fit had negative ξ, bGEV was employed with different values of the quantile $a$ ranging from 0.975 to 0.75, with $b$ set to $a - 0.01$. If the fitted ξ was positive, $a = 0.05, b = 0.2$ were retained from \cite{castro_camilo2022}.

\begin{figure}[htbp]
	\begin{center}
		\includegraphics[width=12cm]{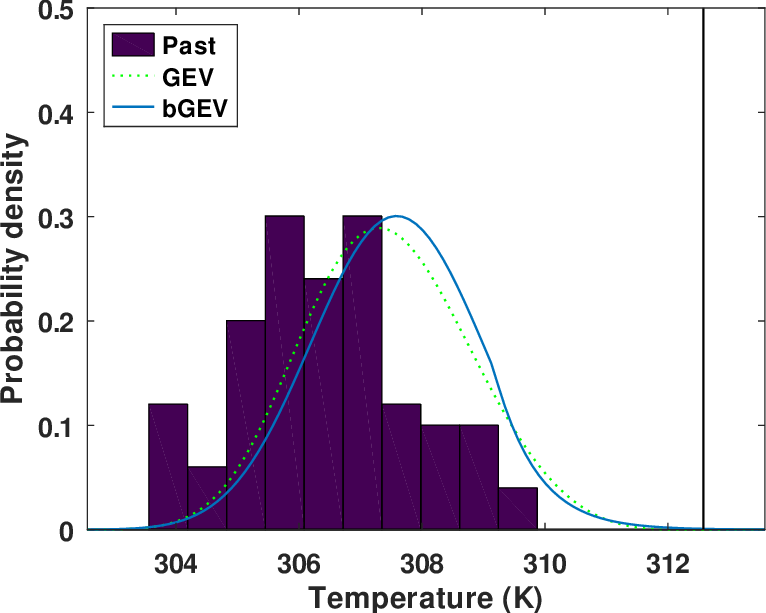}
	\end{center}
	\caption{Example of a new temperature record from near Adelaide, Australia in December 2019, showing the histogram of the 1940-2018 annual maximum temperatures and the forecasts for 2019 based on the GEV and bGEV. Both the GEV and bGEV appropriately shift the forecast probability distribution upward relative to the historic one due to global warming, but the GEV distribution, with negative shape parameter, has zero probability mass past 312.3 K, whereas the bGEV forecast has positive, though small, probability for the actual 312.6 K (vertical line).}
	\label{fig:example_fcst} 
\end{figure}

\begin{figure}[htbp]
	\begin{center}
		\includegraphics[width=12cm]{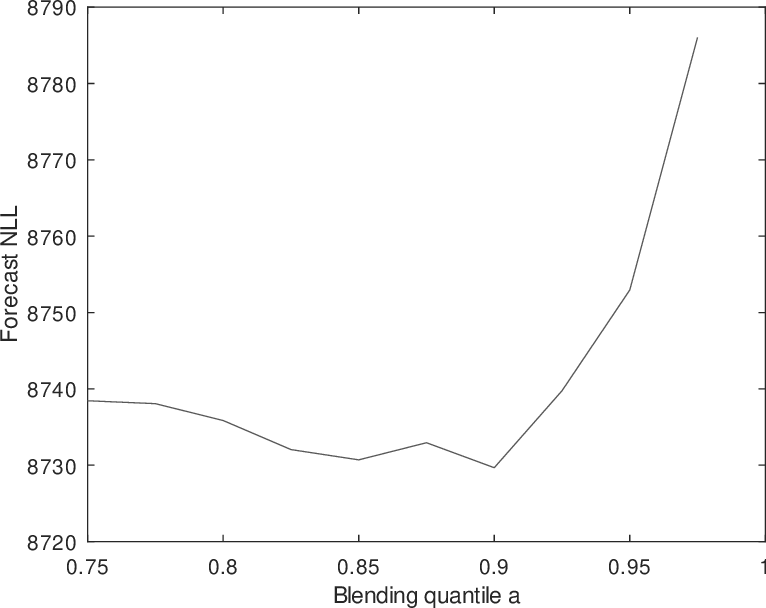}
	\end{center}
	\caption{Summed negative log likelihood (NLL) across 5400 annual maximum temperature forecasts using the bGEV with different blending quantile hyperparameter values $a$. Lower NLL characterizes a better probabilistic forecast. Summed NLL for the different hyperparameter values shown ranged from 8730 to 8786. By comparison, the summed NLL for forecasts using the GEV was infinite, while for forecasts using the Gumbel distribution (GEV with shape parameter fixed at 0) it was 9132.}
	\label{fig:a_NLL} 
\end{figure}

Results showed that for both GEV and bGEV (regardless of $a, b$), most (91\%) of the fitted values of ξ were negative, consistent with previous analyses of heat extremes \cite{frias12, rai2024}, with a median of about $-0.22$. The median $μ_t$ was around 1.4, meaning that annual maximum temperatures warmed some 40\% faster than global mean temperature. For GEV fits, the summed forecast NLL was infinite, because in a few cases (33 out of 5400; example shown in Figure \ref{fig:example_fcst}) the forecast-year temperature hit a new record that was outside the support of the fitted GEV with negative ξ. The bGEV eliminated this problem, so the log likelihoods were all finite regardless of the chosen negative-ξ blending quantiles $a, b$. NLL summed across the 5400 forecasts reached a broad optimum for $a$ between about 0.82 and 0.90 (Figure \ref{fig:a_NLL}). Setting $b$ to $a-0.05$ or $a-0.15$, instead of $a-0.01$, was also tried and gave similar but slightly worse (higher) summed NLL. Overall, these initial results suggest that forecast quality using the bGEV distribution is not very sensitive to the exact choice of blending quantiles.

\begin{figure}[htbp]
	\begin{center}
		\includegraphics[width=12cm]{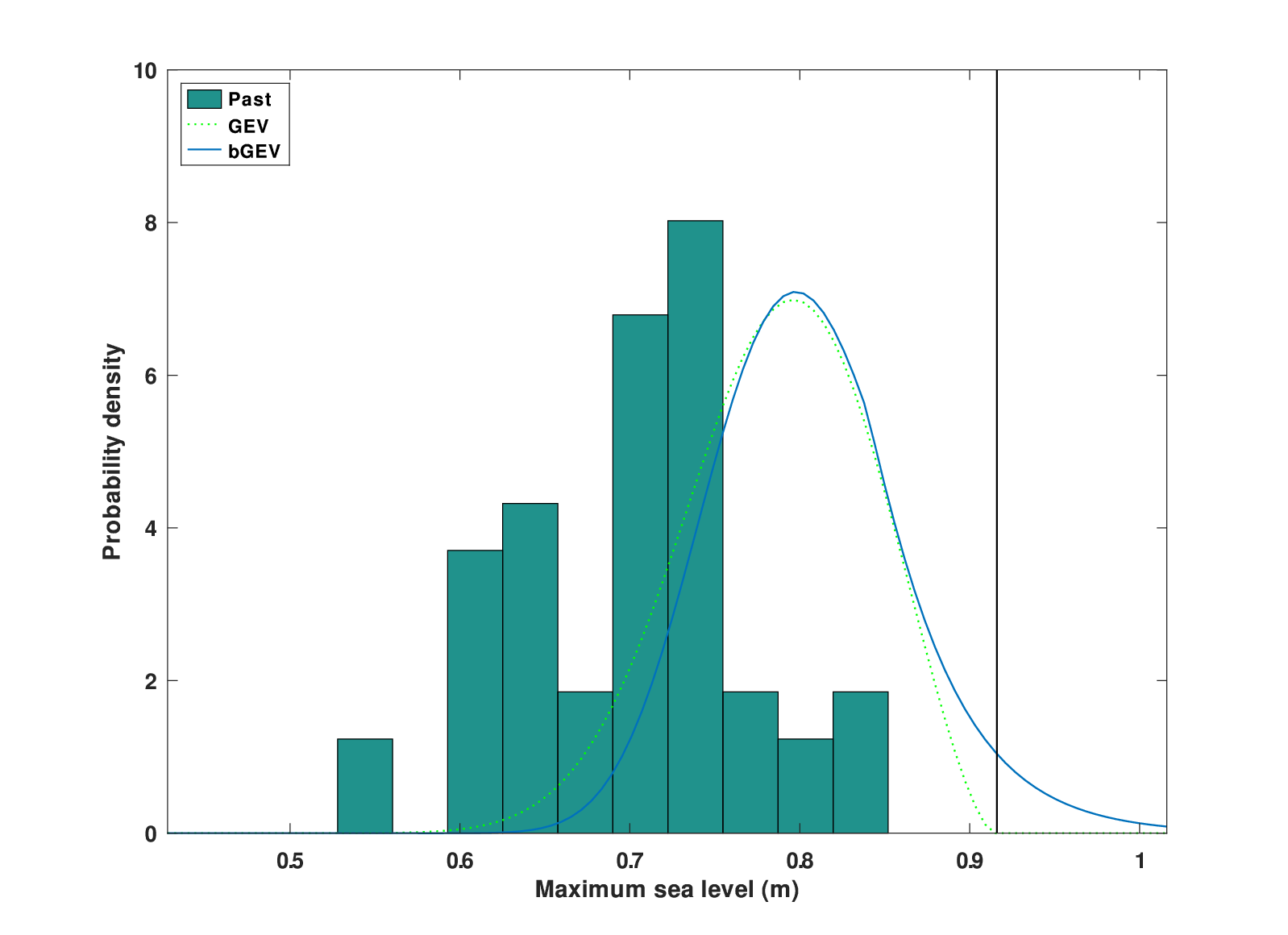}
	\end{center}
	\caption{Example of a new sea level record from Hilo, Hawai'i in 2020, showing the histogram of the 1970-2019 annual maximum sea level and the forecasts for 2020 based on the GEV and bGEV. Both the GEV and bGEV appropriately shift the forecast probability distribution upward relative to the historic one due to global warming and have almost the same forecast mode, but the GEV distribution, with negative shape parameter, has zero probability mass past 0.9143 m, whereas the bGEV forecast has positive probability for the actual 0.916 m (vertical line).}
	\label{fig:example_fcst_sl} 
\end{figure}

As another example application, the GEV and bGEV were compared for one-year-ahead forecasts of annual maximum sea level at Hilo, Hawai'i, for which data was available for 1970-2023 from the National Oceanic and Atmospheric Administration's National Ocean Service. Again allowing at least 30 years of data for fitting, forecasts were successively made for 2001-2023. The location parameter was again taken to be a linear function of global mean temperature. The GEV fit consistently had a negative shape parameter, and the new record sea level set during a storm on January 12, 2020 was higher than the maximum allowable value of the forecast probability distribution, resulting again in an infinite mean forecast NLL. By contrast, the bGEV, again with negative shape parameter, forecast a finite probability for the 2020 record, which was not even necessarily particularly extreme (around the 96th percentile for $a = 0.75, b = 0.74$, which had the lowest summed NLL of the $a, b$ values tried; Figure \ref{fig:example_fcst_sl}). 

\section{Discussion and conclusion}

Many directions could be explored for improving forecasts of extreme events. Partial pooling of information on the bGEV distribution parameters across grid cells using hierarchical Bayesian modeling \cite{gelman12, rahillmarier2022} could enable more precise inference, as could the inclusion of climate covariates besides global mean temperature \cite{belkhiri2021}. Such forecasts could also be used for other variables such as humid heat \cite{lu2023}, and for compound hazards characterized by the simultaneous occurrence of multiple extremes, such as heat and drought or flood \cite{gu2022, yin2022} or compound floods from the co-occurrence of heavy precipitation and coastal storm surge \cite{tanir2021}. A Bayesian approach also allows to include uncertainty due to bGEV parameter fitting in the forecast probability distribution \cite{ragno2019}. Other extensions of the GEV distribution have also been proposed and can be compared with the bGEV distribution, such as transmuted GEV \cite{otiniano2019} and odd GEV \cite{gyasi2023} distributions, as could other distributions unrelated to the GEV that might be more appropriate in specific cases.

In summary, extending the bGEV distribution to cases with negative shape parameter allows better probabilistic forecasts of phenomena such as heat extremes by eliminating the sharp upper bound that is intrinsic to the GEV distribution with negative shape parameter but is not realistic for finite sample lengths. 

\begin{sloppypar}
Octave \cite{eaton12} programs to compute quantities related to the extended bGEV distribution and to fit the bGEV to the simulation and example data described above are available at \url{https://github.com/nir-krakauer/bgev_octave}. These build on functionality for working with the GEV distribution previously implemented in the Octave Statistics package \cite{statistics}.
\end{sloppypar}

\bibliographystyle{unsrtnat}
\bibliography{nir}

 \end{document}